
\input panda
%
\loadamsmath
%
\pageno=0\baselineskip=14pt
\nopagenumbers{
\line{\hfill SWAT/94-95/53}
\line{\hfill US-FT/18-94}
\line{\hfill SNUCTP-94-119}
\line{\hfill\tt hep-th/9412062}
\line{\hfill November 1994}
\ifdoublepage \bjump\bjump\bjump\bjump\else\vfill\fi
\centerline{\capstwo Massive integrable soliton theories}
\bjump
\centerline{\scaps Timothy J. Hollowood$^1$, J. Luis Miramontes$^2$
and Q-Han Park$^3$}
\sjump
\centerline{\sl $^1$Department of Physics, University of Wales Swansea,}
\centerline{\sl Singleton Park, Swansea SA2 8PP, U.K.}
\centerline{\tt t.hollowood@swansea.ac.uk}
\sjump
\centerline{\sl $^2$Departamento de F\'\i sica de Part\'\i culas,}
\centerline{\sl Facultad de F\'\i sica,}
\centerline{\sl Universidad de Santiago,}
\centerline{\sl E-15706 Santiago de Compostela, Spain}
\centerline{\tt miramontes@gaes.usc.es}
\sjump
\centerline{\sl $^3$Department of Physics and}
\centerline{\sl Research Institute for Basic Science,}
\centerline{\sl Kyunghee University,}
\centerline{\sl Seoul, 130-701, Korea}
\centerline{\tt qpark@nms.kyunghee.ac.kr}
\bjump
\ifdoublepage
\vfill
\eject\null\vfill\fi
\centerline{\capsone ABSTRACT}\sjump
Massive integrable field theories in $1+1$ dimensions are defined at
the Lagrangian level,
whose classical equations of motion are related to
the ``non-abelian'' Toda field
equations. They can be thought of as generalizations of the
sine-Gordon and complex sine-Gordon theories.
The fields of the theories take values in a non-abelian Lie group and
it is argued that the coupling constant is quantized, unlike the
situation in the sine-Gordon theory, which is a special case since its field
takes values in an abelian group. It is further shown that these
theories correspond to perturbations of certain coset conformal field
theories. The solitons in the theories will, in general, carry
non-abelian charges.
\sjump\vfill
\ifdoublepage \else
\fi
\eject}
\yespagenumbers\pageno=1
%
%

\chapter{Introduction}

In this letter we construct new series of integrable field theories in $1+1$
dimensions. We show that the simplest examples are the
sine-Gordon and complex sine-Gordon theories, respectively.
The former is
well-known to be integrable at the quantum level and an exact
factorizable S-matrix has been found
for the scattering of the classical
lump solutions; in this case topological solitons and breathers [\Ref{SG}].
Remarkably, the fundamental
field of the theory is the ground-state of the breather
system, so that all states in the theory can be understood from a
semi-classical analysis of the lump solutions.
Recently, the complex sine-Gordon theory has been quantized
semi-classically and on the basis of this an exact factorizable
S-matrix was proposed to describe the scattering of the lumps of the
theory [\Ref{DH}] (following earlier work [\Ref{DM}]);
in this case charged solitons. Like the sine-Gordon theory the
fundamental field of the theory corresponds to one of classical lump
solutions. These two theories serve as paradigms of what can be
expected for the more complex situations: we expect the lump-like
classical solutions to admit a semi-classical quantization from which
the exact factorizable S-matrix could be deduced. In this
sense the paper should be regarded as the first stage in a programme
to find these exact S-matrices. The theories in
general have a non-abelian global symmetry (the complex sine-Gordon is
simpler since its symmetry group is abelian); so the solitons will
carry non-abelian charges (as well as topological charges in some
cases). In this sense they are similar to the Skyrme model in four
dimensions (see for example [\Ref{SKY}]).

The classical
integrable equations that underlie the theories are the ``non-abelian
Toda equations'' of Leznov and Saveliev [\Ref{LS}]. What was not so
clear from their original work---and is the subject of the present
paper---is whether these theories can be written in Lagrangian form
and hence
can be used to define relativistic quantum field theories. It
transpires that many of the these classical equations cannot be
derived from a Lagrangian
with a positive-definite kinetic term and a real potential term;
however, there are several families for which this can be done.
Most of the theories which admit a Lagrangian formalism, having
a positive-definite kinetic energy and real potential energy,
have massless degrees-of-freedom because the potential energy has flat
directions. Whilst these theories may be interesting in their own
right, they will not admit a factorizable S-matrix (at least in the
conventional sense); however, we show how the massless
degrees-of-freedom may be removed to give a purely massive
integrable field theory.

In contrast with the usual Toda field theories, where the field takes
values in the Cartan subgroup of a Lie group, the field
$h(x,t)$ of the theories
we shall discuss, takes values in a non-abelian Lie group $G_0$. The kinetic
term of the theory is simply the WZW action for the group $G_0$, so
that the actions of the theories are of the form
$$
S[h]={1\over\beta^2}\left\{S_{\rm WZW}[h]-\int d^2x\,V(h)\right\},
\nfr{LAG}
where $V(h)$ is some potential function on the group manifold,
to be specified, and $\beta$ is a
coupling constant which plays no role in the classical theory.\note{So
in these theories, just as in the sine-Gordon theory, the
semi-classical limit is the same as the weak-coupling limit.} The
potential has a minimum which can be chosen to be at the identity
$h=1$, and so expanding around the minimum by taking $h=1+i\phi+\cdots$,
one can see that the quadratic term in the kinetic energy is
$$
{1\over8\pi\beta^2}\int d^2x\,{\rm Tr}\left(\partial_\mu\phi
\partial^\mu\phi\right),
\efr
where ${\rm Tr}(\ )$ is the suitably normalized trace in some
faithful representation of the Lie algebra $g_0$ associated to $G_0$.
If the kinetic term is to be positive-definite then
immediately we see that the group $G_0$ must be the compact group,
unless $G_0$ is abelian, in which
case it can be both the compact or maximally non-compact group (in the latter
case with $\beta^2\rightarrow-\beta^2$ so that the kinetic term is
positive definite rather than negative definite).

An important consequence of the form of the Lagrangian
is that if the group $G_0$ is non-abelian (and compact) and the
quantum theory is to be well-defined then the coupling constant has
to be quantized:
$$
\beta^2={1\over k},\qquad k\in{\Bbb N}.
\efr
Such a quantization of the coupling constant does not occur in the
sine-Gordon theory or the Toda theories because the group is abelian
in these cases. So an important consequence of this is that in the
quantum theory there will not be a continuous coupling constant; this
will have important implications for the construction of the exact
S-matrix of the theories. An example of this quantization of the
coupling constant occurs in the complex sine-Gordon theory [\Ref{DH}]
which, as we have mentioned, is the simplest theory of this type
with a non-abelian field.

\chapter{The Leznov and Saveliev construction}

We now turn to the definition of the potential $V(h)$. We will
choose the
potential so that the classical equations of motion are one of the
integrable equations constructed by Leznov and Saveliev [\Ref{LS}]
(the so-called ``non-abelian Toda equations'') and
hence can be written in zero-curvature form.

The Leznov and Saveliev construction starts by considering an
sl(2) embedding of some
finite Lie algebra $g$ specified by the
generators $\{J_\pm,J_0\}$. The Cartan element $J_0$ induces a
gradation of $g$, by adjoint action:
$$
g=\bigoplus_{k=-N}^Ng_k,\qquad [J_0,a]=ka\quad{\rm for}\ a\in g_k.
\nfr{GRAD}
In general, $k$ runs over the half-integers because there are both
integer and half-integer spin representations when $g$ is
decomposed under the sl(2); however, the Leznov and Saveliev
construction in its original form
is restricted to integral embeddings, i.e. there can be no
half-integer spin representations and $k$ runs over the integers only.
(By definition $J_\pm\in g_{\pm1}$.)

The field $h(x,t)$ takes values in the group $G_0$ associated to the
Lie algebra of the zero-graded component $g_0$. The associated
integrable equation is of the form
$$
\partial_-\left(h^{-1}\partial_+h\right)=m^2\left[J_+,h^{-1}J_-h\right],
\nfr{CFT}
where $\partial_\pm=\partial_t\pm\partial_x$ are the light-cone
derivatives and $m$ is a mass scale. This equation can be written in
zero-curvature form as
$$
\left[\partial_++h^{-1}\partial_+h+imJ_+,\partial_--imh^{-1}J_-
h\right]=0.
\efr
In spite of the explicit mass scale,
these equations are classically conformally invariant
and are actually generalizations of the Liouville
theory, which is recovered by taking $g={\rm sl}(2)$ with
$$
h=\pmatrix{e^\phi&0\cr o&e^{-\phi}\cr},\qquad J_+=\pmatrix{0&1\cr
0&0\cr},\qquad J_-=\pmatrix{0&0\cr
1&0\cr}.
\efr
Such conformally invariant theories
with non-compact groups---hence indefinite kinetic
terms---have been studied in the context of two-dimensional black-holes
[\Ref{BH}].

In this paper we wish to define massive field theories. Leznov and
Saveliev showed how to generalize the aforementioned
construction to break conformal
invariance whilst preserving integrability (these theories have also
been discussed in [\Ref{JU}]). The idea involves generalizing
equations \CFT\ to
$$
\partial_-\left(h^{-1}\partial_+h\right)=m^2\left[\Lambda_+,
h^{-1}\Lambda_-h\right],
\nfr{MFT}
where $\Lambda_\pm=J_\pm+Y_\pm$. Integrability is then maintained if
the constant elements $Y_\pm\in g_{\mp N}$, the minimal and maximal
graded components of $g$, respectively. The conformal invariance is
manifestly broken because the elements $\Lambda_\pm$ do not have
definite grade.

The Leznov and Saveliev
construction of integrable equations describing massive fields
has a lot freedom, firstly due to the choice of the sl(2) embedding and
secondly due to choice of the elements $Y_\pm$. However, if we
wish to describe theories that can be written at the Lagrangian level
with a positive definite kinetic energy then we shall find that there
are many fewer
possibilities. First of all, we must choose the form of the group
$G_0$. As we have already pointed out, if $g_0$ is non-abelian,
in order to get a theory with a
positive-definite kinetic term
we are forced to take the compact form of the group $G_0$.
This imposes a reality condition on the field
$h^\dagger=h^{-1}$ which is
consistent with the equations of motion \MFT\
only if $\Lambda^\dagger_\pm=\Lambda_\pm$ and so
$J_\pm^\dagger=Y_\pm$. But since $J_\pm^\dagger\in
g_{\mp1}$ this implies that the sl(2) embedding has to have $N=1$.
However, in general, the resulting theory will not be invariant under
parity. Such theories may be interesting but here we
will study theories invariant under parity, for which we need
$\Lambda_+=\Lambda_-$.\note{This is clear from the equations
of motion which have the
symmetry $x_\pm\mapsto x_\mp$ along with $h\mapsto h^{-1}$, only if
$\Lambda_+=\Lambda_-$.} This requires $Y_\pm=J_\mp$ and so
$J_\pm^\dagger=J_\mp$; hence
$$
\Lambda\equiv\Lambda_+=\Lambda_-=J_++J_-.
\efr

The element $\Lambda$ is actually
an element of a Heisenberg subalgebra of the loop algebra of
$g$. In fact the language of Heisenberg subalgebras, which is more
suited to describing the associated integrable hierarchies of
equations [\Ref{HS}], can be used instead of the language of sl(2)
embedding; however for our purposes we will find it sufficient to
stick to the language of sl(2) embeddings.

To summarize: if $g_0$ is non-abelian then we can only construct a
field theory with a positive-definite kinetic term
if the sl(2) embedding of $g$ induces a gradation of the form
$$
g=g_{-1}\oplus g_0\oplus g_1.
\nfr{LOC}
Hence, under the sl(2), $g$ decomposes into triplets and singlets only.

With these restrictions the equations of motion may be derived by
extremizing the action
$$
S[h]={1\over\beta^2}\left\{S_{\rm WZW}[h]-{m^2\over8\pi}
\int d^2x\,{\rm Tr}\left(h\Lambda h^{-1}\Lambda\right)\right\}.
\nfr{ACTION}
So the Lagrangian has the advertized form \LAG\ with a potential
$$
V(h)={m^2\over8\pi}{\rm Tr}\left(h\Lambda h^{-1}\Lambda\right).
\efr
Notice that since $\Lambda^\dagger=\Lambda$ the potential is
real when $h$ takes values in the compact group. The action \ACTION\ can be
obtained by Hamiltonian reduction of the two loop WZW model associated
to the affine untwisted Kac-Moody algebra $g^{(1)}$~[\Ref{NOS}]; this
generalizes the well known relation between the reduced WZW model and
the non-abelian conformal Toda models~[\Ref{DUBLIN}].

If $g_0$ is abelian then the above restriction to the
compact group does not apply because $G_0$ can then be chosen to be
either the compact or maximally non-compact group, corresponding to
$h^\dagger=h^{-1}$ and $h^\dagger=h$, respectively. The abelian case
arises when the sl(2) embedding is the principal embedding in $g$.
In this case
$$
J_+=\sum_{j=1}^r E_{\alpha_j},
\efr
a sum over the step generators corresponding to the simple roots of
$g$, and $g_{-N}$ is spanned by the step generator $E_{\alpha_0}$,
corresponding to the lowest root. Therefore,
$$
\Lambda_+=\sum_{j=0}^rE_{\alpha_j}.
\efr
So with $G_0$ being the compact group we see that the equations of
motion are not consistent with the reality condition except when
$g={\rm su}(2)$ when
the resulting theory is the sine-Gordon theory. On the other hand,
if we
choose the abelian group $G_0$ to be the maximally non-compact group
then the equations of motion \MFT\ are real by virtue of the fact that
$h^\dagger=h$ and $\Lambda_+^\dagger=\Lambda_-$. These theories are the
affine Toda field equations associated to an algebra $g$.

Returning to the non-abelian theories, we see that the potential has a
minimum when $h=1$, the identity in $G_0$. If we expand the action
around the minimum $h=\exp i\phi$, with $\phi^\dagger=\phi$, for small
$\phi$, we have
$$
S={1\over8\pi\beta^2}\int d^2x\,\left\{{\rm Tr}\left(\partial_\mu\phi
\partial^\mu\phi
\right)-{m^2\over2}{\rm Tr}\left(
[\phi,\Lambda][\phi,\Lambda]\right)+\cdots\right\},
\efr
up to a constant.\note{The WZ term is a total derivative
at the quadratic order and hence does not contribute at this order.}
So we expect the quantum theory to contain a set of particles with
masses found by diagonalizing these quadratic terms. It is
straightforward to see that there will be massless particles
associated to the subalgebra $g_0^0\subset g_0$ of elements
which commute with
$\Lambda$. So the potential, therefore, will have flat
directions and the theory will have a mixture of
massless and massive degrees-of-freedom. Such theories may be
interesting to study in there own right, since they describe
renormalization group trajectories which interpolate between the
WZW models based on $G_0$ and $G_0^0$ (the compact group associated to
$g_0^0$).

However, in this paper we wish to study theories with a
mass gap, because such theories will admit an S-matrix description.
To achieve this, we have to somehow introduce the constraints
that $P(h^{-1}\partial_+h)=0$ and $P(h\partial_-h^{-1})=0$,
where $P$ is the projection operator onto the
subalgebra $g_0^0\subset g_0$.
The way to introduce this constraint was discussed
in [\Ref{QHI}]. First of all, notice that the potential is invariant
under the
diagonal group action $h\mapsto\alpha h\alpha^{-1}$, for $\alpha\in
G_0^0$. So the idea is to {\it gauge\/} this diagonal $G_0^0$ group
action by
introducing a gauge field $A_\pm$ taking values in $g_0^0$. The kinetic
term is then the gauged WZW action $S_{\rm WZW}[h,A_\pm]$
which describes the $G_0/G_0^0$ coset conformal field
theory [\Ref{GWZW}]. The action of the theory is then
$$
S[h,A_\pm]={1\over\beta^2}\left\{S_{\rm WZW}[h,A_\pm]-
{m^2\over8\pi}\int d^2x\,{\rm Tr}\left(h\Lambda h^{-1}
\Lambda\right)\right\}.
\nfr{MLAG}
The action of the gauged WZW model is explicitly [\Ref{GWZW}]
$$\eqalign{
&S_{\rm WZW}[h,A_\pm]=S_{\rm WZW}[h]\cr
&+{1\over2\pi}\int d^2x\,{\rm
Tr}\left(A_+h\partial_-h^{-1}+A_-h^{-1}\partial_+h+A_+hA_-h^{-1}-
A_+A_-\right).\cr}
\efr
The variation of \MLAG\ with respect to $h$ gives the equation of motion
which can be written in the zero-curvature form
$$
\left[\partial_++h^{-1}\partial_+h+h^{-1}A_+h+imz\Lambda,
\partial_-+A_--imz^{-1}h^{-1}\Lambda h\right]=0,
\nfr{EQA}
where we have introduced the spectral parameter $z$ whose inclusion
plays an important role in establishing the integrability of the equations.
Variations with respect to the gauge field leads to the constraints
$$
\eqalign{
P\left(h\partial_-h^{-1}+hA_-h^{-1}\right)-A_-&=0,\cr
P\left(h^{-1}\partial_+h+h^{-1}A_+h\right)-A_+&=0,\cr}
\nfr{CON}
where $P$ is, as before, the projector onto the subalgebra $g_0^0$.
By projecting \EQA\ onto $g_0$ one can see that the gauge field is
flat: $[\partial_++A_+,\partial_-+A_-]=0$ which reflects the vector
gauge invariance of the action.

To show that the constraints \CON\ remove the massless
degrees-of-freedom it suffices to choose the gauge $A_+=A_-=0$, which
is consist due to the flatness of the gauge field and the vector gauge
invariance of the action. In this gauge
the equations of motion \EQA\ reduce to
$$
\partial_-\left(h^{-1}\partial_+h\right)=m^2
\left[\Lambda,h^{-1}\Lambda h\right],
\efr
along with the constraints \CON:
$$
P\left(h^{-1}\partial_+ h\right)=0,\qquad P\left(h\partial_-h^{-1}\right)=0.
\efr
As pointed out in [\Ref{QHI}], the constraints cannot be
solved locally in this gauge;
however, there are different gauge choices
for which the constraint equations can be solved in terms of local
fields.

So the theories that we end up with can be considered as integrable
perturbations of certain coset conformal field theories.
Perturbations of these conformal field theories have been considered
recently in [\Ref{QHI},\Ref{IB}]. In the latter reference,
it was shown how a perturbation of a $G_0/G_0^0$ coset model of the
form ${\rm Tr}(
hTh^{-1}\bar T)$ was classically integrable if the constant elements
$T,\bar T\in g_0$ lie in the centralizer
of $g_0^0$. Notice that this construction
is rather similar to the one discussed here, except that in our models
$\Lambda$, the analogue of $T$, lies outside $g_0$ in the larger
algebra $g$. However, in section 4 we shall find some overlap between
these two formalisms.

Notice that the theories have a global $G_0^0$ symmetry. This is
because the potential in \MLAG, and therefore the action itself,
is invariant under both left and right action by $G_0^0$:
$h\rightarrow\alpha h$ and $h\rightarrow h\alpha$, $\alpha\in G_0^0$.
Therefore after gauging the diagonal subgroup there is a residual
$G_0^0$ symmetry.

The conformal dimension of the perturbing operator can be found by
taking an operator product expansion with the stress tensor of the
coset model. This highlights an important property of these
perturbations: in general the dimension of the perturbing operator
depends upon the
representation chosen for $g$ for defining the potential. So the
formalism potentially generates a number of inequivalent integrable
perturbations of the coset model.

\chapter{${\rm sl}(2)$ embeddings with $N=1$ and the models}

In the appendix we explain how to derive all the sl(2) embeddings
with $N=1$. In this section we put these results together
with the formalism of the last section to write down integrable
perturbations of various coset models.

Notice from the form of the action \ACTION, and the invariance of the
Haar measure in the path integral, that theories related by the
fact that $\Lambda_1$ and $\Lambda_2$ are conjugate, so that
$\Lambda_1=\alpha\Lambda_2\alpha^{-1}$ for a constant element
$\alpha\in G_0$, are
actually identical since they are related by a transformation of the field.
This means that we need only consider sl(2)
embeddings up to conjugation by the group $G_0$. In addition, we are
restricting ourselves to the parity invariant theories, so that
$J_\pm^\dagger=J_\mp$.

For the
classical Lie algebras we shall use the orthonormal vectors $\{e_j\}$
and the dual basis $\{e_j^\vee\}$, with $e_j^\vee(e_k)=\delta_{jk}$, in
terms of which one can describe both the root space and the Cartan
subalgebra of $g$, respectively.

For $A_1$ the principal embedding itself has $N=1$. In the two-dimensional
representation
$$
J_0={1\over2}\pmatrix{1&0\cr 0&-1\cr},\qquad\Lambda=\pmatrix{0&1\cr
1&0\cr}.
\nfr{AO}
The subalgebra $g_0$ consists of the Cartan subalgebra and $g_0^0=\emptyset$.
In this case the field is valued in U(1) and in the next section we
show that it is the sine-Gordon theory.

For general $A_{n-1}$, $n$ has to be even, $n=2p$, and there is an $N=1$
embedding, up to conjugation, for which
$$
J_0={1\over2}\sum_{j=1}^p\left(e_j^\vee-e_{p+j}^\vee\right),\qquad
J_+=\sum_{j=1}^pE_{e_j-e_{p+j}}.
\efr
The compact form of the algebra is su($2p$) where an element of the
algebra is a traceless hermitian matrix.\note{In our convention the
generators of the compact Lie algebra will be hermitian.} In the defining
$2p$-dimensional representation we may take
$$
J_0={1\over2}\pmatrix{1_p&0\cr 0&-1_p\cr},\qquad\Lambda=\pmatrix{0&1_p\cr
1_p&0\cr}.
\nfr{ANJ}
So the subalgebra $g_0$ consists of elements of the form
$$
\pmatrix{a&0\cr 0&b\cr},\qquad a^\dagger=a,\ b^\dagger=b,\
{\rm Tr}(a)+{\rm Tr}(b)=0;
\efr
hence $g_0={\rm su}(p)\oplus{\rm su}(p)\oplus{\rm u}(1)$. The
subalgebra $g_0^0$
consists of elements which commute with $\Lambda$ and thus of
the form
$$
\pmatrix{a&0\cr 0&a\cr},\qquad a^\dagger=a,\ {\rm Tr}(a)=0;
\efr
hence $g_0^0={\rm su}(p)$: the diagonal embedding.
So the model corresponds to an integrable perturbation of the coset
$$
{{\rm SU}(p)\times{\rm SU}(p)\over{\rm SU}(p)}\times{\rm U}(1) .
\efr
An element of the group $G_0$ can be written as
$$
h=\pmatrix{h_1&0\cr 0&h_2\cr},\qquad h_1,h_2\in{\rm U}(p),\ {\rm
det}(h_1){\rm det}(h_2)=1.
\efr
For this particular representation the potential has the form
$$
V(h_1,h_2)={m^2\over8\pi}{\rm Tr}\left(h_1h_2^{-1}+h_2h_1^{-1}\right).
\efr
The case when $p=2$ is discussed more fully in the next section.

For $C_n$ there is only one embedding with $N=1$, up to
conjugation; namely
$$
J_0={1\over2}\sum_{j=1}^ne_j^\vee,\qquad
J_+=\sum_{j=1}^nE_{2e_j},
\nfr{CONE}
which follows from (A.9) with $p=0$ and $r=n$.
The compact form of the algebra is sp($n$)\note{In our notation
sp($n$) has rank $n$.} where an element of the
algebra in the defining $2n$-dimensional representation has the block
form
$$
\pmatrix{a&b\cr b^\star&-a^\star\cr},\qquad a^\dagger=a,\ b^T=b,
\efr
where $a$ and $b$ are $n$ dimensional matrices.
In this representation we can take
$$
J_0={1\over2}\pmatrix{1_n&0\cr 0&-1_n\cr},\qquad
\Lambda=\pmatrix{0&1_n\cr 1_n&0\cr}.
\efr
The subalgebra $g_0$ consist of elements of the form
$$
\pmatrix{a&0\cr 0&-a^\star},\qquad a^\dagger=a;
\nfr{GOCN}
hence $g_0={\rm su}(n)\oplus{\rm u}(1)$. The elements which commute
with $\Lambda$ are of the form
$$
\pmatrix{a&0\cr 0&a\cr},\qquad a^\star=-a,\ a^T=-a,
\efr
and so $g_0^0={\rm so}(n)$. So the model corresponds to a perturbation of
the coset
$$
{{\rm U}(n)\over{\rm SO}(n)}.
\efr
In this representation, an element of $G_0$ has the form
$$
h=\pmatrix{\tilde h&0\cr 0&\tilde h^\star\cr},\qquad\tilde h\in{\rm U}(n),
\nfr{OTRO}
and the potential is
$$
V(\tilde h)={m^2\over8\pi}
{\rm Tr}\left(\tilde h\tilde h^T+\tilde h^\star\tilde h^{-1}\right).
\efr

We now turn to the algebras of the orthogonal groups.
For both $B_r$ and $D_r$ there is an $N=1$ embedding with
$$
J_0=e_1^\vee,\qquad
J_+=E_{e_1-e_2}+E_{e_1+e_2},
\nfr{ONEE}
where we have realized the embedding in terms of a regular subalgebra
$D_2\subset B_r\; {\rm or}\; D_r$. The compact form of the algebras
is so($n$) (corresponding to
$B_r$ if
$n=2r+1$ and $D_r$ if $n=2r$). The defining
representation consists of purely imaginary antisymmetric $n$-dimensional
matrices. In this representation we can take
$$
J_0={i\over2}\pmatrix{0&1&\cdots\cr -1&0&\cdots\cr \vdots&\vdots&\ddots\cr},
\qquad\Lambda=i\pmatrix{0&0&0&\cdots\cr 0&0&1&\cdots\cr
0&-1&0&\cdots\cr \vdots&\vdots&\vdots&\ddots\cr},
\efr
where the elements not shown are zero. The subalgebra $g_0$
consists of matrices of the form
$$
i\pmatrix{0&a&\cdots\cr -a&0&\cdots\cr \vdots&\vdots&b\cr},
\efr
where $a$ is a real number and $b$ is a real $(n-2)$-dimensional
antisymmetric matrix;
hence $g_0={\rm so}(n-2)\oplus{\rm u}(1)$. The elements which
commute with $\Lambda$ are those of the form
$$
i\pmatrix{0&0&0&\cdots\cr 0&0&0&\cdots\cr 0&0&0&\cdots\cr
\vdots&\vdots&\vdots&c\cr},
\efr
where $c$ is a real $(n-3)$-dimensional antisymmetric matrix; therefore
$g_0^0={\rm so}(n-3)$. So the model
corresponds to an integrable perturbation of the coset
$$
{{\rm SO}(n-2)\over{\rm SO}(n-3)}\times{\rm U}(1).
\efr
In this representation the field has the block form
$$
h=\pmatrix{A&0\cr 0&\tilde h\cr},\qquad\tilde h\in{\rm SO}(n-2),
\efr
where the U(1) factor is
$$
A=\exp\pmatrix{0&a\cr -a&0\cr}.
\efr
In this case the potential cannot be written down in a neat way.
The models admit a reduction, preserving integrability,
which involves restricting the field to
be an element of the subgroup SO$(n-2)\subset{\rm SO}(n-2)\times{\rm U}(1)$.
To see that this is consistent
with the equations of motion, one only has to notice that
$[\Lambda,h^{-1}\Lambda h]\in{\rm SO}(n-2)$ when $A=1$. Hence, the
reduced theory will describe an integrable deformation of the coset
SO$(n-2)/{\rm SO}(n-3)\sim{\rm S}^{n-3}$.
The example with $n=5$ is considered more
explicitly in the next section.

For $D_{2p}$, where $p$ is an integer, there is an additional $N=1$ embedding
specified by
$$
J_0={1\over2}\sum_{j=1}^{2p}e_j^\vee,\qquad
J_+=\sum_{j=1}^pE_{e_{2j-1}+e_{2j}}.
\nfr{EB}
The compact form of the algebra is so($4p$). In the $4p$-dimensional
defining representation, elements of the algebra correspond to purely
imaginary antisymmetric matrices. In this representation we can take, in
block form,
$$
J_0={i\over2}\pmatrix{0&1_{2p}\cr -1_{2p}&0\cr},\qquad\Lambda=
i\pmatrix{j&0\cr 0&-j\cr},
\efr
where the $2p$-dimensional matrix $j$ is in block form
$$
j=\pmatrix{0&1_p\cr-1_p&0\cr}.
\efr
Elements of the subalgebra $g_0$ have the following block form
$$
i\pmatrix{a&b\cr-b&a\cr},\qquad a^T=-a,\ b^T=b,
\efr
where both $a$ and $b$ are real $2p$-dimensional matrices. We can
amalgamate these to form $2p$-dimensional
hermitian matrix $b+ia$, showing that $g_0={\rm
su}(2p)\oplus{\rm u}(1)$. The elements of $g_0$ which commute with
$\Lambda$ are those for which $[a,j]=0$ and $\{b,j\}=0$. These
conditions can be written
$$
j(b+ia)=-(b+ia)^Tj,
\efr
which is the defining relation for the Lie algebra sp($p$). Hence the
model corresponds to a perturbation of the coset
$$
{{\rm U}(2p)\over{\rm Sp}(p)}.
\efr

For $D_{2p}$ there is an additional non-conjugate sl(2) embedding
got by replacing $e_{2p}\mapsto-e_{2p}$ in \EB. However, although the
embedding is non-conjugate it is related by a diagram symmetry which
means that the resulting theory is equivalent.

For the exceptional Lie algebras, there is only one sl(2)
embedding with $N=1$, which is in $E_7$. In this case the model will
correspond to some integrable perturbation of a coset of $E_6\times{\rm
U}(1)$, although it is not very instructive to write down the
potential in this case.

Notice that for all these embeddings $g_0$ has a U(1) factor. This
factor can be traced to the generator $J_0\in g_0$ which commutes with
the rest of $g_0$ by construction and hence always forms a u(1) subalgebra.
As we have seen and discuss more fully in the next section, for some
of the models this U(1) field can be decoupled whilst preserving
integrability.

\chapter{Some explicit examples}

In this section we will consider more explicitly some of the theories
defined above.

The first example we consider is the principal embedding in $A_1$. In
the two-dimensional representation of \AO\ the field is
$$
h=\pmatrix{e^{i\phi}&0\cr 0&e^{-i\phi}\cr},
\efr
and the action can be written explicitly in terms of the field $\phi$
as
$$
S[\phi]={1\over8\pi\beta^2}\int d^2x\,\left((\partial_\mu\phi)^2-m^2\cos2\phi
\right).
\efr
This example is nothing but the ubiquitous sine-Gordon theory. In
this case the WZ-term does not contribute because the group is
abelian, and so correspondingly there will be no quantization of the coupling
constant in the quantum theory. The spectrum of the model consists of
solitons and breathers whose exact S-matrix was
written down in [\Ref{SG}]. Notice that all the states of the theory
appear as the quantization of classical lump solutions.

The second example we consider, and in some respects the simplest
after the sine-Gordon theory,
since $g_0$ has the smallest rank,
is the theory based on $B_2$ or $C_2$. In fact it is easier to work with
$C_2$ since its vector representation has dimension 4
(corresponding to the spinor representation of $B_2$). There is one
embedding, up to conjugation, which in $C_2$ we can take to be
$$
J_+=E_{e_1+e_2},
\efr
this being conjugate to \CONE. So in 2$\times$2 block form
$$
\Lambda=J_++J_-=\pmatrix{0&\sigma_1\cr \sigma_1&0\cr}.
\efr
where $\sigma_1$ is one of the Pauli matrices:
$$
\sigma_1=\pmatrix{0&1\cr 1&0\cr},\ \sigma_2=\pmatrix{0&i\cr -i&0\cr},\
\sigma_3=\pmatrix{1&0\cr 0&-1\cr}.
\efr
In the same block form, the field $h\in G_0$ is from \OTRO
$$
h=\pmatrix{\tilde h&0\cr 0&\tilde h^\star\cr},
\efr
where $\tilde h$ is an element of the group U(2) ($\simeq{\rm
SO}(3)\times{\rm U}(1)$) in the two-dimensional
representation. The group $G_0^0$ consists of elements in the Cartan
subalgebra of the SU(2) factor; hence, elements of the form
$$
\pmatrix{e^{ia}&0&0&0\cr 0&e^{-ia}&0&0\cr 0&0&e^{-ia}&0\cr 0&0&0&e^{ia}\cr},
\efr
for arbitrary $a$. The equations of motion \EQA\ decouple into two
equations involving the $2\times2$ matrix $\tilde h$, which are complex
conjugate to each other. Before gauge fixing one of the equations is
$$
\left[\partial_++\tilde h^{-1}\partial_+\tilde h+\tilde h^{-1}A_+\tilde h,
\partial_-+A_-\right]+m^2\left(\sigma_1\tilde h^T\sigma_1\tilde h-
\tilde h^{-1}\sigma_1\tilde h^\star\sigma_1\right)=0,
\nfr{BGF}
where the gauge fields $A_\pm$ take values in the Cartan subalgebra of
the SU(2), i.e. its components are
proportional to $\sigma_3$.

As we mentioned in the last section, the equations of
motion admit a reduction---preserving
integrability---by restricting
$\tilde h\in{\rm SU}(2)$ so that the reduced theory is actually
a perturbation of the coset SU$(2)/$U(1).
In this case we will write the equations of motion in terms of local
fields which means that we must choose a gauge different from
$A_\pm=0$. One way to fix the gauge is to choose a parameterization of
the form (as in [\Ref{QHI}])
$$
\tilde h=\pmatrix{u&i\sqrt{1-uu^\star}\cr
i\sqrt{1-uu^\star}&u^\star\cr}.
\efr
The constraints \CON\ can then be solved for the gauge field:
$$
A_+={u^\star\partial_+u-u\partial_+u^\star\over4(1-uu^\star)}\sigma_3,
\qquad A_-={u\partial_-u^\star-u^\star\partial_-u
\over4(1-uu^\star)}\sigma_3.
\efr
The equations of motion in this gauge are then from \BGF
$$
\partial_-\partial_+u+{u^\star\partial_+u\partial_-u
\over1-uu^\star}+m^2u(1-uu^\star)=0,
\efr
and its complex conjugate. This is precisely the complex sine-Gordon
equation. The fact that this theory could be derived from perturbing a
coset model was first pointed out in [\Ref{IB}] and discussed further
in [\Ref{QHI}]. What we have found is that the theory fits quite
naturally into the class of non-abelian Toda theories, a fact which
was not apparent in [\Ref{QHI}].

The theory exhibits charged soliton solutions [\Ref{CSG},\Ref{DM}],
where the charge corresponds to the residual
U(1) symmetry $u\rightarrow e^{i\theta}u$. Furthermore it
has been argued that the theory is
integrable at the quantum level [\Ref{DH},\Ref{DM}]. The
theory corresponds to an integrable perturbation of the SU(2)/U(1)
coset model and in this case the dimension of the perturbing operator
is $\Delta=\bar\Delta=2/(k+2)$, independent of the representation
chosen for the potential. The solitons have
internal motion which is indicative of the existence of the global
U(1) symmetry and rather surprisingly the
particle states of the theory are actually identified with particular
solitons (in much the same way that the particle of the sine-Gordon
theory corresponds to a breathing solution). The full S-matrix of
the theory has been found in [\Ref{DH}].

The final set of models
we consider are those associated to su($2p$). Let us take the vector
representation in which $\Lambda$ is given in \ANJ.
An element of the group $G_0={\rm SU}(p)\times{\rm SU}(p)\times{\rm
U}(1)$ is of the form
$$
h=\pmatrix{h_1&0\cr 0&h_2\cr},
\efr
where $h_1,h_2\in{\rm U}(p)$, with det$(h_1){\rm det}(h_2)=1$.
The equations of motion in the $A_\pm=0$ gauge
are
$$
\partial_-(h_1^{-1}\partial_+h_1)=m^2\left(h^{-1}_2h_1-h_1^{-1}h_2
\right),\qquad h_1^{-1}\partial_+h_1+h_2^{-1}\partial_+h_2=0.
\nfr{EXO}

When $p=2$, as for the theory associated to so(5), the
theory admits an integrable reduction
by setting det$(h_1)={\rm det}(h_2)=1$, which is clearly
consistent with \EXO. In this case the theory describes an integrable
deformation of the coset model SU$(2)\times{\rm SU}(2)/{\rm SU}(2)$.
This is precisely
the model discussed in [\Ref{QHII}]. The conformal dimension of the
perturbation in this case is  $\Delta=\bar\Delta=3/(4+2k)$. These
equations have soliton solutions [\Ref{QHIII}] and it would be
interesting to proceed to a semi-classical quantization and hence
determine the exact S-matrix.
In these theories the perturbation does depend on the
representation chosen for $g_0$. So if
we choose $h_1$ and $h_2$ to be in the representation of spin $j$ then
one can show that the perturbation of the coset model has
dimension $\Delta=\bar\Delta=2j(j+1)/(k+2)$.

\chapter{Discussion}

We have constructed $1+1$ dimensional field theories which are
classically integrable. The classical equations of motion are related
to the non-abelian Toda equations and these equations admit soliton
solutions [\Ref{JU}]. These theories are naturally to be thought of as
generalizations of the sine-Gordon and complex sine-Gordon theories.
So we expect that, in common with those theories, the spectrum of
quantum states can be understood in terms of the semi-classical
quantization of various lump-like solutions (solitons and breathers).
The theories in general have a global non-abelian symmetry group
and so we
expect that the soliton solutions will transform in representations of
the symmetry group.
We pointed out an essential difference between the theories whose
fields take values in a group whose algebras are non-abelian
and the sine-Gordon theory, where the group is abelian:
in the former the coupling constant is
quantized at the quantum level.

We have shown that these theories can be viewed as integrable
perturbations of certain coset conformal field theories and using
this picture should allow one to establish whether integrability
survives at the quantum
level, via the method of Zamolodchikov [\Ref{INTZ}].

We have seen that in certain cases it was possible to perform a
reduction of the models, whilst maintaining integrability. In fact
there are a number of possibilities for making such reductions. In
general, the idea is that if there exists a subalgebra $\tilde g\subset
g_0$, with corresponding compact group $\tilde G\subset G_0$, and
$[\Lambda,\tilde h^{-1}\Lambda\tilde h]\in\tilde g$ for $\tilde h\in\tilde
G$, then the reduction of the model to $\tilde G$ will be integrable.
So as well as the reductions already mentioned, there are many other
possibilities; two examples are
$$
{{\rm SU}(p)\times{\rm SU}(p)\over{\rm SU}(p)}\times{\rm U}(1)
\rightarrow{{\rm SO}(p)\times{\rm SO}(p)\over{\rm SO}(p)},
\efr
and
$$
{{\rm SU}(2q)\times{\rm SU}(2q)\over{\rm SU}(2q)}\times{\rm U}(1)
\rightarrow{{\rm Sp}(q)\times{\rm Sp}(q)\over{\rm Sp}(q)}.
\efr

The outstanding problem is to find the classical spectrum of solitons
and breathers of these theories and then perform a semi-classical
quantization. From this it should be possible to infer the exact
S-matrices.

\jump
TJH would like to thank Ioannis Bakas for explaining his work and
Jonathan Evans for useful conversations. JLM thanks Luiz Ferreira for
his enlightening comments. QP is supported partly by KOSEF, Kyunghee
Univ. and BSRI-94-2442. JLM is supported partially by CICYT (AEN93-0729)
and DGICYT (PB90-0772). TJH is supported by a PPARC Advanced Fellowship.

\appendix{}
%

The sl(2) embeddings $\{J_\pm, J_0\}$ of simple Lie algebras
have been classified by Dynkin~[\Ref{DYN}]; $J_0$ is called the {\it
defining vector} and it characterizes the embedding up to conjugation
by an automorphism of
$g$. It is always possible to choose a system of simple roots of the
algebra $\{\alpha_1,\ldots , \alpha_r\}$ such that the numbers $s_i =
\alpha_i(J_0)$ are $0$, $1/2$, or $1$. These numbers are associated to
the nodes of the Dynkin diagram of $g$, and the diagram, with the
numbers written on, is called the {\it characteristic} of the
embedding. A necessary and sufficient condition for two sl(2)
subalgebras to be conjugate is that their characteristics coincide.
Moreover, a necessary and sufficient condition that the embedding be
integral is that all the $s_i$'s are either $0$ or $1$. We will
express the characteristic as a vector ${\bf s}=(s_1,\ldots, s_r)$,
and  we shall use that particular choice of the system of simple roots
of $g$ from now on.

Let us restrict ourselves to integral embeddings and consider a
positive root of $g$
$$
\alpha = \sum_{i=1}^{r} \> n_i \> \alpha_i\>, {\rm \;\;\; with\;\;\;}
n_i \in {\Bbb Z}\geq0 {\rm \;\;\;for\ all\;\;\;} i=1,\ldots, r\>;
\efr
then
$$
[J_0, E_{\pm\alpha}] \> =\> \pm \left(\sum_{i=1}^{r} n_i\> s_i
\right)\>  E_{\pm\alpha}\>.
\efr
Therefore, if
$$
\theta = \sum_{i=1}^{r} \> k_i\> \alpha_i\>
\efr
is the highest root of $g$, where $\{k_1,\ldots, k_r\}$ are the Kac
labels of $g$, then, in eq. \GRAD, $E_{\pm\theta}\in g_{\pm N}$ with
$$
N \>= \> \sum_{i=1}^{r} k_i\> s_i\>.
\efr
Consequently, $N=1$ if, and only if, the characteristic of the
embedding is such that $s_i= \delta_{ij}$ with $k_j=1$.

Moreover,
let $i_1,\ldots,i_p$ be all the indices for which $s_{i_1}=\cdots
=s_{i_p}=0$, then, the zero-graded subalgebra $g_0$ is
isomorphic to a direct sum of the $(r-p)$-dimensional centre and the
semisimple Lie algebra whose Dynkin diagram is the subdiagram of the
Dynkin diagram of $g$ consisting of the nodes $i_1,\ldots,i_p$. We
can now list all the sl(2) embeddings of a simple Lie algebra which
correspond to gradations having $N=1$.

For the exceptional Lie algebras, their Kac labels
show that the only possibilities are either $E_6$, with ${\bf s}
=(1,0,0,0,0,0)$ or ${\bf s} =(0,0,0,0,1,0)$, or $E_7$, with ${\bf s}
=(0,0,0,0,0,1,0)$.  The characteristics of the different sl(2)
embeddings for these Lie algebras can be found in
[\Ref{DYN}]\note{In [\Ref{DYN}], the characteristic is
normalized to be $2{\bf s}$ and its components equal $0$, $1$, or
$2$ ($0$ or $2$ if the embedding is integral).}, and the only embedding
with the required characteristic corresponds to $E_7$ with ${\bf s}
=(0,0,0,0,0,1,0)$, which is associated to a regular subalgebra
$3A_1\subset E_7$, and whose corresponding zero-graded subalgebra is
$$
\left( E_7\right)_0 = E_6 \oplus {\rm u}(1)\>.
\efr
Therefore, we conclude that this is the only sl(2) embedding with
$N=1$ of the exceptional Lie algebras $G_2$, $F_4$, $E_6$, $E_7$, and
$E_8$.

For the classical Lie algebras, the
sl(2) subalgebras can be realized as the principal sl(2)
subalgebra of a regular subalgebra of $g$, up to a few exceptions
occurring in $D_n$---for a review, see, for example [\Ref{SORB}],
where very detailed expressions for the defining
vectors of the embeddings are provided.

For $A_n$, as all the Kac labels equal one, $s_i=\delta_{ij}$ for some
$j=1,\ldots,n$; but, using the results of~[\Ref{SORB}] it can be
checked that there is an embedding with this characteristic only when
$n$ is odd, $n=2p-1$, and $j=p$. Then
$$
{\bf s}=(\> \underbrace{0,\ldots,0\>}_{p-1\;\; {\rm times}}\>,
1,\underbrace{0,\ldots,0\>}_{p-1\;\; {\rm times}}\>)\>,
\efr
and it is associated to the regular subalgebra
$$
\underbrace{A_1 \oplus \cdots \oplus A_1}_{ p\;\; {\rm times}}\>\subset\>
A_{2p-1}\>.
\efr
In this case, the zero-graded subalgebra is
$$
\left(A_{2p-1}\right)_0 = A_{p-1}\oplus A_{p-1}\oplus {\rm u}(1)\>.
\efr

For $C_n$, the only possibility allowed by the Kac
labels is ${\bf s}=(0,\ldots,0,1)$, and there is only
one embedding with this characteristic. It can be realized in terms of
any of the following regular subalgebras
$$
\underbrace{A_1 \oplus \cdots \oplus A_1}_{p\;\; {\rm times}} \>\oplus\>
\underbrace{C_1 \oplus \cdots \oplus C_1}_{r\;\; {\rm times}}\>\subset
\>C_n\>,
\efr
where $p$ and $r$ are arbitrary non-negative integers such that
$n=2p+r$, and $C_1$  is a regular $A_1$ subalgebra of $C_n$
corresponding to a long root; all these different realizations
are conjugate by inner automorphisms of $C_n$. The zero-graded
subalgebra is
$$
\left(C_n\right)_0 = A_{n-1} \oplus {\rm u}(1)\>.
\efr

For $B_n$, the value of the Kac labels leads to ${\bf s} =
(1,0,\ldots,0)$, and there is only one embedding with this
characteristic. It can be realized in terms of any of the two regular
subalgebras
$$
B_1 \subset B_n \quad {\rm and}\quad D_2\subset B_n\>,
\efr
where $B_1$ is a regular $A_1$ subalgebra associated to a short root,
and $D_2$ is a regular $A_1\oplus A_1$ subalgebra with
$J_+ = E_{e_j-e_{j+1}}\> +\> E_{e_j+e_{j+1}}$ for some $j=1,\ldots,
n-1$; these two realizations are also conjugate by an inner
automorphism of $B_n$. The zero-graded
subalgebra is
$$
\left(B_n\right)_0 = B_{n-1}\oplus {\rm u}(1)\>.
\efr

Finally, for $D_n$, the Kac labels admit three possibilities. The
first one is ${\bf s}=(1,0,\ldots,0)$, and there is an sl(2)
embedding with this characteristic associated to the regular subalgebra
$$
D_2\subset D_n\>.
\efr
In this case, the zero-graded subalgebra is
$$
\left( D_n\right)_0 = D_{n-1}\oplus {\rm u}(1)\>.
\efr
The other two possibilities arise when $n$ is even, $n=2p$; they
are
$$
{\bf s}=(0,\ldots,0,1) \quad {\rm and}\quad {\bf s}'=(0,\ldots,1,0)\>,
\efr
and there are two embeddings associated to two non-conjugate
regular subalgebras of the form
$$
\underbrace{A_1 \oplus \cdots \oplus A_1}_{ p\;\; {\rm times}}\>
\subset\> D_{2p}\>,
\efr
whose corresponding zero-graded subalgebras are
$$
\left( D_{2p}\right)_0 = A_{2p-1}\oplus {\rm u}(1)\>.
\efr
These last two sl(2) embeddings are conjugate by the
diagram automorphism of $D_{2p}$ that takes $\alpha_{2p-1}
\leftrightarrow \alpha_{2p}$, but not by any inner automorphism of
$D_{2p}$. Finally, let us mention that for $D_4$ the three embeddings
are conjugate by diagram automorphisms.

\references

\beginref
\Rref{DYN}{E.B.~Dynkin, Amer. Math. Soc., Transl. Ser. 2, {\bf 6}
(1957) 111-244.}
\Rref{SORB}{L.~Frappat, E.~Ragoucy and P.~Sorba, Commun. Math. Phys.
{\bf 157} (1993) 499-548.}
\Rref{DH}{N. Dorey and T.J. Hollowood, {\sl Quantum scattering of
charged solitons in the complex sine-Gordon theory\/},
Preprint SWAT/46, {\tt hep-th/9410140}}
\Rref{QHI}{Q-H. Park, Phys. Lett. {\bf B328} (1994) 329}
\Rref{QHII}{Q-H. Park and H.J. Shin, {\sl Deformed minimal models and
generalized Toda theory\/}, Preprint SNUCTP-94-83, {\tt
hep-th/9408167}}
\Rref{IB}{I. Bakas, Int. J. Mod. Phys. {\bf A9} (1994) 3443}
\Rref{GWZW}{D. Karabali, Q-H. Park, H.J. Schnitzer and Z. Yang, Phys.
Lett. {\bf B216} (1989) 307\newline
D. Karabali and H.J. Schnitzer, Nucl. Phys. {\bf B329} (1990)
649\newline K. Gawedski and A. Kupiainen, Phys. Lett. {\bf B215}
(1989) 119; Nucl. Phys. {\bf B320} (1989) 625}
\Rref{LS}{A.N. Leznov and M.V. Saveliev, Commun. Math. Phys. {\bf 89}
(1983) 59}
\Rref{JU}{D.I. Olive, M.V. Saveliev and J. Underwood, Phys. Lett. {\bf
B311} (1993) 177\newline
J. Underwood, {\sl Aspects of non-abelian Toda theories\/},
 {\tt hep-th/9304156}}
\Rref{QHIII}{Q-H. Park and H.J. Shin, in preparation}
\Rref{BH}{J-L. Gervais and M.V. Saveliev, Phys. Lett. {\bf B286} (1992)
271}
\Rref{DM}{H.J. de Vega and J.M. Maillet, Phys. Lett. {\bf B101} (1981)
302; Phys. Rev {\bf D28} (1983) 1441}
\Rref{CSG}{B.S. Getmanov, JETP Lett. {\bf25} (1977) 119}
\Rref{SG}{A.B. Zamolodchikov and Al.B. Zamolodchikov, Ann. Phys.
{\bf120} (1979) 253}
\Rref{INTZ}{A.B. Zamolodchikov, Int. J. Mod. Phys. {\bf A3} (1988)
743}
\Rref{HS}{M.F. de Groot, T.J. Hollowood and J.L. Miramontes, Commun.
Math. Phys. {\bf145} (1992) 57}
\Rref{SKY}{T.H.R. Skyrme, Proc. Roy. Soc. {\bf A260} (1961)
127\newline G. Adkins, C. Nappi and E. Witten, Nucl. Phys. {\bf B228}
(1983) 552}
\Rref{NOS}{H.~Aratyn, L.A.~Ferreira, J.F.~Gomes and A.H.~Zimerman, Phys. Lett.
{\bf B254} (1991) 372\newline
L.A.~Ferreira, J.L.~Miramontes and J.~S\'anchez Guill\'en,
{\sl Solitons, Tau-functions and Hamiltonian reduction for non-abelian
affine Toda theories\/}, Preprint US-FT/21-94}
\Rref{DUBLIN}{J.~Balog, L.~Feh\'er, P.~Forgacs, L.~O'Raifeartaigh and
A.~Wipf, Phys. Lett. {\bf B227} (1989) 214; Ann. of Phys. {\bf 203}
(1990) 76\newline
L.~Feher, L.~O'Raifeartaigh, P.~Ruelle, and I.~Tsutsui, Ann. of Phys. {\bf 213}
(1992) 1}
\endref
\ciao